\documentclass[11pt]{article}
\usepackage{moriond,epsfig}

\def\Journal#1#2#3#4{{#1} {\bf #2}, #3 (#4)}

\def\PRD{{\em Phys. Rev.} D}

\def\be{\begin{equation}}
\def\ee{\end{equation}}
\def\bea{\begin{eqnarray}}
\def\eea{\end{eqnarray}}

\begin{document}
\vspace*{4cm}
\title{EXTRACTING INFORMATION ABOUT WIMP PROPERTIES FROM DIRECT DETECTION EXPERIMENTS: ASTROPHYSICAL UNCERTAINTIES}

\author{ ANNE M.  GREEN}

\address{School of Physics and Astronomy, University of Nottingham, University Park, Nottingham, NG7 2RD, UK}

\maketitle\abstracts{WIMP direct detection experiments probe the ultra-local dark matter density and velocity distribution. We review how uncertainties in these quantities affect the accuracy with which the WIMP mass and cross-section can be constrained or determined.}

\section{Introduction}

If Weakly Interacting Massive Particles (WIMPs) are detected then measuring the WIMP mass and cross-section will shed light on the particle nature of the WIMP. It will allow us to constrain models of particle physics beyond the standard model, for instance supersymmetry and universal extra dimensions. Furthermore a convincing WIMP discovery may require
consistent signals (i.e. same inferred mass) in multiple experiments in multiple channels (direct, indirect, collider). We discuss the effect of astrophysical uncertainties on the accuracy with which the WIMP
mass and cross-section could be measured from direct detection experiments.

Direct detection experiments aim to detect WIMPs in the Milky Way halo via the energy deposited when they elastically scatter off detector nuclei~\cite{ddtheory,book}. Assuming spin-independent coupling the differential event rate (number of events per unit energy, 
per unit time, per unit detector mass) is given by~\cite{jkg,ls}
\be
\label{drde}
\frac{{\rm d} R}{{\rm d}E} =
             \frac{\sigma_{{\rm p}} 
             \rho_{\chi}}{2 \mu_{{\rm p} \chi}^2 m_{\chi}}
             A^2 F^2(E)   \int^{\infty}_{v_{{\rm min}}} 
            \frac{f(v)}{v} {\rm d}v     \,, 
\ee
where $\rho_{\chi}$ is the ultra-local WIMP density, $f(v)$ the ultra-local WIMP speed distribution in the rest frame of the detector, $\sigma_{{\rm
p}}$ the WIMP scattering cross section on the proton, $\mu_{{\rm p} \chi} = 
(m_{\rm p} m_{\chi})/(m_{{\rm p}}+ m_{{\chi}})$ the WIMP-proton reduced mass, 
$A$ and $F(E)$ the mass number and
form factor of the target nuclei respectively and $E$ is the recoil
energy.
The lower limit of the integral, $v_{{\rm min}}$, is the minimum
WIMP speed that can cause a recoil of energy $E$:
\begin{equation}
\label{vmin}
v_{{\rm min}}=\left( \frac{ E m_{A}}{2 \mu_{{\rm A} \chi}^2} 
             \right)^{1/2} \,,
\end{equation}
where $m_{A}$ is the atomic mass of the detector nuclei
and $\mu_{{\rm A} \chi}$ the WIMP-nucleon reduced mass.

The energy, time and direction dependence of the event rate provide potential signals for 
discriminating WIMP events from backgrounds.
If we assume, as a starting point, the standard halo model with an isotropic Gaussian speed distribution
\be
\label{max}
f(v) \propto \exp{ \left( - \frac{v^2}{v_{\rm c}^2} \right)} \,,
\ee
where $v_{\rm c}$ is the local circular speed (the speed with which stars in the solar neighbourhood orbit the Galactic centre),
 then the differential event rate has a characteristic energy~\cite{ls}
\be
\label{er}
E_{\rm R} =  \frac{2 \mu_{{\rm A} \chi}^2 v_{\rm c}^2}{m_{\rm A}} \,.
\ee
If $m_{\chi} \ll m_{\rm A}$ then $E_{\rm R} \propto m_{\chi}^2$, while if $m_{\chi} \gg m_{\rm A}$ then $E_{\rm R} \sim {\rm const}$. Therefore, provided the WIMP mass is not too large or not small~\footnote{If the WIMP mass is too small the event rate above the detector energy threshold becomes small.}, a measurement of the WIMP mass can be extracted from the energy spectrum of events.

The motion of the Earth with respect to the Galactic rest frame provides two potential WIMP `smoking guns'. Due to the Earth's orbit about the Sun the net velocity of the lab with respect to the Galactic rest frame varies over the course of a year. The net speed is largest in the Summer and hence there are more  high speed WIMPs, and less low speed WIMPs, in the lab frame.
This produces an energy dependent, ${\cal O}(5\%)$, annual modulation in the
differential event rate~\cite{dfs}. To detect this a large exposure (detector mass times time) is required.

Our motion with respect to the Galactic rest frame also
produces a directional signal~\cite{spergel}. The WIMP flux in the lab frame is
sharply peaked in the direction of motion of the Sun (towards the constellation CYGNUS).  The recoil spectrum is then peaked in the direction opposite to this. This is a large signal; with an ideal detector only 10 events would be needed to demonstrate that the recoil rate was isotropic~\cite{mgs} and $\sim 30$ to confirm that the median recoil direction is opposite to the direction of solar motion~\cite{billard,gm}. A detector which can measure the recoil directions is required though (e.g. DMTPC, DRIFT, MIMAC, NEWAGE). See Ref.~\cite{ahlen} for a review of the current status of directional detection experiments.

\section{Astrophysical input}
Direct detection experiments probe the ultra-local dark matter velocity and density distribution on sub milli-pc scales. They are in fact the only probe of the dark matter distribution on such scales. Numerical
simulations and astronomical simulations can only resolve $\sim 100\, {\rm pc}$ scales, many orders of magnitude larger.

The standard halo model usually used in the analysis of data from direct detection experiments is an isothermal sphere with an isotropic Maxwellian velocity distribution, as given in eq.~(\ref{max}), and a local density $\rho_{0} = 0.3 \, {\rm GeV} \, {\rm cm}^{-3}$. The dark matter halos formed in numerical simulations are far more complicated than this, however. They are anisotropic, not perfectly spherical and contain substructure. The standard halo model is therefore likely to be a poor approximation to the Milky Way halo.

\subsubsection{Numerical simulations}
In the past few years several groups have carried out high resolution simulations of the formation
of Milky Way like halos, in a cosmological context~\cite{hansen,fs,vogelsberger,kuhlen}.
They typically find  speed distributions which deviate systematically from a multivariate Gaussian (the simplest anisotropic generalisation of the Maxwellian distribution). There are more low speed particles, and the peak in the distribution is lower. There are also broad stochastic features at high speeds. The detailed properties of these features vary from halo to halo and are thought to reflect the formation history of the halo~\cite{vogelsberger}.

Simulations can't tell us the value of the ultra-local dark matter density, however they do provide insight into how the local (on $\sim$ kpc scales) density at the Solar radius varies. While simulated halos contain large amounts of substructure, the inner regions are fairly smooth and the variation in density at the Solar radius~\footnote{Halos are not perfectly spherical, so the averaging is in fact done over an ellipsoidal shell.} is fairly small, $<15\%$~\cite{vogelsberger}.

There are two caveats to these results. Firstly, as mentioned above, the scales resolved by simulations
are many orders of magnitude larger than those probed by direct detection experiments. 
Kamionkowski and Koushiappas have used simulation inspired models to calculate the probability distribution of the ultra-local dark matter density~\cite{kk}. The find distributions which are skew positive and peak at densities lower then the mean density (due to substructure). The detailed properties of the probability distributions (e.g. the most likely ultra-local density) depend on the assumptions/extrapolations made.
Vogelsberger and White have developed a new technique to study the ultra-local dark matter distribution~\cite{vw}. They find that the ultra-local dark matter consists of a huge number of streams and is essentially smooth. This suggests that the ultra-local dark matter density and velocity distribution should not be drastically different to those on the scales resolved by simulations.

Secondly the simulations discussed above contain dark matter only, while baryons dominate in the inner regions of the Milky Way. Simulating baryonic physics is extremely difficult, and producing galaxies whose detailed properties match those of real galaxies is an outstanding challenge. Some recent simulations have found that late merging sub-halos are preferentially dragged towards the disc, where they are destroyed leading to the formation of a rotating dark disc~\cite{read1,read2,ling}.

\subsubsection{Observations}
Dark matter halos can not be observed directly, however their structure can be probed using visible tracers.

The local dark matter density has traditionally~\cite{co,ggt} been thought to lie in the range $\rho_{0} \sim 0.2-0.8 \, {\rm GeV} \, {\rm cm}^{-3}$, i.e. with a factor of a few uncertainty. Recent determinations have
significantly smaller statistical errors.
Widrow et al.~\cite{widrow}, using spherical halo models with a cusp ($\rho(r) \propto r^{-\alpha}$ as $r \rightarrow 0$),
find $\rho_{0} \sim (0.3 \pm 0.05) \, {\rm GeV} \, {\rm cm}^{-3}$. Catena and Ullio~\cite{cu}, using NFW and Einasto profiles, find  $\rho_{0} \sim (0.39 \pm 0.03) \, {\rm GeV} \, {\rm cm}^{-3}$. On the other hand
Weber and de Boer~\cite{weber}, using a wider range of models for the dark matter density profile, have found values in the range $\rho_{0} \sim 0.2-0.8 \, {\rm GeV} \, {\rm cm}^{-3}$, and Salucci et al.~\cite{salucci}
find $\rho_{0} \sim 0.42 \pm 0.15 \, {\rm GeV} \, {\rm cm}^{-3}$ using a DM density profile independent method. In summary, while the statistical errors are now at the ${\cal O}(10\%)$ level, the systematic errors that arise from uncertainties in modeling the Milky Way are clearly still larger than this.

The local velocity dispersion is related to the local circular speed~\footnote{For the standard halo model the 3-d velocity dispersion is $\sigma = \sqrt{3/2} \, v_{\rm c}$, in general there is a relationship between the radial velocity dispersion, $\sigma_{\rm r}$, and $v_{\rm c}$ which depends on the density profile and the velocity anisotropy, e.g. Ref.~\cite{bt}.}, 
which is traditionally taken to be
$v_{c} = (220 \pm 20 ) \, {\rm km \, s}^{-1}$~\cite{kerr}. Recent determinations span a wide range
($(254 \pm 16 ) \, {\rm km \, s}^{-1}$~\cite{reid}, $(200 - 280 ) \, {\rm km \, s}^{-1}$~\cite{mcmillan}, $(236 \pm 11 ) \, {\rm km \, s}^{-1}$~\cite{bovy}),  which suggests that modeling uncertainties are significant here too.

\section{Implications}
\subsubsection{Constraints on measurements of $\sigma_{\rm p}$}
The normalization of the event rate is proportional to the product of the cross-section, $\sigma_{\rm p}$, and the local density, $\rho$, therefore the uncertainties in $\rho$ translate directly into uncertainties in $\sigma_{\rm p}$. In this case the uncertainty is the same for exclusion limits (or in the case of a detection, measurements) from different experiments.

\subsubsection{Time averaged differential event rate}
For most current experiments, which measure the time averaged differential event rate, the most important uncertainty is that in the circular velocity, $v_{\rm c}$. The characteristic energy, $E_{\rm R}$, depends on both the WIMP mass and $v_{\rm c}$. By differentiating eq.~(\ref{er}) one can see that the uncertainty in $v_{\rm c}$ leads to an uncertainty in the WIMP mass:
\be
\frac{\Delta m_{\chi}}{m_{\chi}} = \left[ 1 + \left( \frac{m_{\chi}}{m_{A}} \right) \right] \frac{\Delta v_{\rm c}}{v_{\rm c}} \,.
\ee

The effect of uncertainties in the {\em shape} of the speed distribution is typically relatively small. This is because the energy spectrum is proportional to an integral over the speed distribution, and hence changes in the shape are effectively smoothed out. An exception to this is if the experiment is only sensitive to the high speed tail of the speed distribution, which is the case if the WIMP is light or the threshold energy is high. This is because the uncertainties in the tail of the speed distribution are larger,
and the smoothing effect is smaller.  Changing the shape of the speed distribution leads to a, typically small, change in the shape of exclusion limits. The change depends on the experiment, as different experiments probe different WIMP speed ranges (depending on the energy threshold and the mass of the target nuclei)~\cite{greenexclude}.
In the case of a detection, the shape of the speed distribution leads to a few per-cent systematic uncertainty in the accuracy with which $m_{\chi}$ could be measured from the energy spectrum~\cite{greenmass}. With multiple detectors (with different target nuclei) $m_{\chi}$ could in principle be measured without any assumptions about the speed distribution~\cite{drees}.

\subsubsection{Annual modulation}

As the annual modulation arises from the small shift in the lab speed distribution between Summer and Winter it is far more sensitive to the detailed velocity distribution. This has been studied by numerous authors. The amplitude, phase and even shape of the modulation can change, and the region of WIMP mass-cross-section parameter space corresponding to the DAMA annual modulation signal changes
significantly (e.g. Ref.~\cite{dama}).

\subsubsection{Direction dependence}

The main directional signal, the rear-front asymmetry of the event rate, is robust to the uncertainties in the speed distribution~\cite{mgs}.  However the detailed direction dependence can vary, for instance
the peak recoil direction of high energy recoils may deviate somewhat from the direction of solar motion~\cite{kuhlen}. In fact with sufficient directional data it would be possible to reconstruct the WIMP velocity distribution~\cite{gondolo}.

\section{Summary and future directions}

The energy, time and direction dependence of the direct detection event rate all depend on the ultra-local dark matter distribution. Deviations from the standard halo model which affect these signals are expected. The uncertainty in the ultra-local dark matter density translates directly into an uncertainty in the event rate and hence measurements of (or constraints on) the WIMP cross-section. The uncertainty in the circular speed (which is related to the WIMP velocity dispersion) leads to an
uncertainty in characteristic scale of energy spectrum and hence measurements of the WIMP mass.
The uncertainty in the shape of WIMP velocity dispersion produces an uncertainty in
the amplitude and phase of the annual modulation signal and hence the WIMP
parameters.

Several approaches to dealing with the impact of astrophysical uncertainties have recently been proposed. Strigari and Trotta~\cite{strigari} have suggested using data (such as measurements of the kinematics of Milky Way halo stars and the local escape speed) and a model for the Milky Way mass distribution in a Monte Carlo Markov Chain analysis of direct detection data. Peter~\cite{peter} has 
presented an approach which involves combining data sets from different direct detection experiments and jointly constraining  a parametrisation of the WIMP speed distribution and
the WIMP parameters (mass and cross-section). Both of these approaches are contingent on formulating a reliable, robust model of the Milky Way.

In the long term the dependence of the direct detection event rate on the ultra-local dark matter
distribution is an opportunity rather than an inconvenience. With sufficient data direct detection experiments (in particular directional ones) could measure the ultra-local dark matter
distribution and shed light on the formation of the Milky Way.

\section*{Acknowledgments}
The author is supported by STFC.

\end{document}